\documentclass[reprint,amsmath,amssymb,aps,prl]{revtex4-1}
\usepackage[utf8]{inputenc}
\usepackage{graphicx}
\usepackage{dcolumn}
\usepackage{bm}
\usepackage[space]{grffile}
\usepackage[export]{adjustbox} 
\usepackage{booktabs}
\usepackage[caption=false]{subfig}
\usepackage{natbib}

\begin{document}

\title{Memory in nonmonotonic stress relaxation of a granular system}
\author{Kieran A. Murphy, Jonathon W. Kruppe, Heinrich M. Jaeger}
\affiliation{James Franck Institute, The University of Chicago, Chicago, IL 60637, USA}
\date{\today}

\begin{abstract}
We demonstrate experimentally that a granular packing of glass spheres is capable of storing memory of multiple strain states in the dynamic process of stress relaxation.  Modeling the system as a non-interacting population of relaxing elements, we find that the functional form of the predicted relaxation requires a quantitative correction which grows in severity with each additional memory and is suggestive of interactions between elements.  Our findings have implications for the broad class of soft matter systems that display memory and anomalous relaxation.
\end{abstract}

\maketitle

Subjected to a perturbation, many  systems in nature will relax anomalously (i.e., non-exponentially) over long timescales, suggesting complex dynamics and common underlying, far-from-equilibrium physics.  Examples of specifically logarithmic relaxation are  the slow breaching of colloidal particles at an interface \cite{LogBreachingManoharan2011},  the magnetization decay in type-II superconductors \cite{SuperconductorLog1993},  the dynamics of crumpled elastic sheets \cite{WittenNagelCrumple2002,Lahini2017},  the evolving area of frictional contact between two interfaces \cite{FinebergFrictional2010,Dillavou2018}, and the stress decays in a granular packing at fixed strain \cite{MakseRelaxation2005,LogAgingAcoustic2013}. 

A versatile framework applied to viscoelastic \cite{GenMaxwellModel1989} and dielectric materials \cite{FeldmanDRTReview2002,GuoGuoDRT1983,JonscherReviewDRT199} idealizes a relaxing system as an ensemble of simple, exponential relaxers in parallel with one another, with a distribution of different relaxation times (DRT).  To explain the widespread occurrence of logarithmic relaxation, Amir et al. \cite{Imry2012} motivated a specific distribution of relaxation times on fairly general grounds.  Importantly, the Amir, Oreg, and Imry (AOI) variant of DRT can also explain nonmonotonic relaxation observed in crumpled mylar \cite{Lahini2017}, a frictional interface \cite{Dillavou2018}, and bulk rock salt \cite{HeRockSalt2019}, after subjecting such systems to a particular driving protocol.  

Nonmonotonic relaxation is a surprising and non-intuitive phenomenon.  In the process of energy dissipation, with no external input after the initial driving, a state variable evolves in one direction before turning around after some timescale that was imprinted during the prior driving history. In contrast with memories that are revealed only when the system is driven \cite{KeimNagel2019}, these memories reside in dynamic processes and thus offer a foothold into studying the quixotic march to equilibrium of a far-from-equilibrium system.

Here we employ a granular packing to study nonmonotonic relaxation within the AOI DRT framework.  We find the capacity for memories should theoretically be larger than two, and store the memory of an additional strain state in experiment by appending a compression step at the end of a two-step driving protocol.  The functional form of the relaxation,
while qualitatively similar to that predicted by the model, requires a quantitative correction which grows in severity with the additional memory.  We suggest a route to reconciliation between the model and experiment, guided by the presence of discrete relaxation events in the data.

\begin{figure*}
	\centering
	\includegraphics[width=1\linewidth]{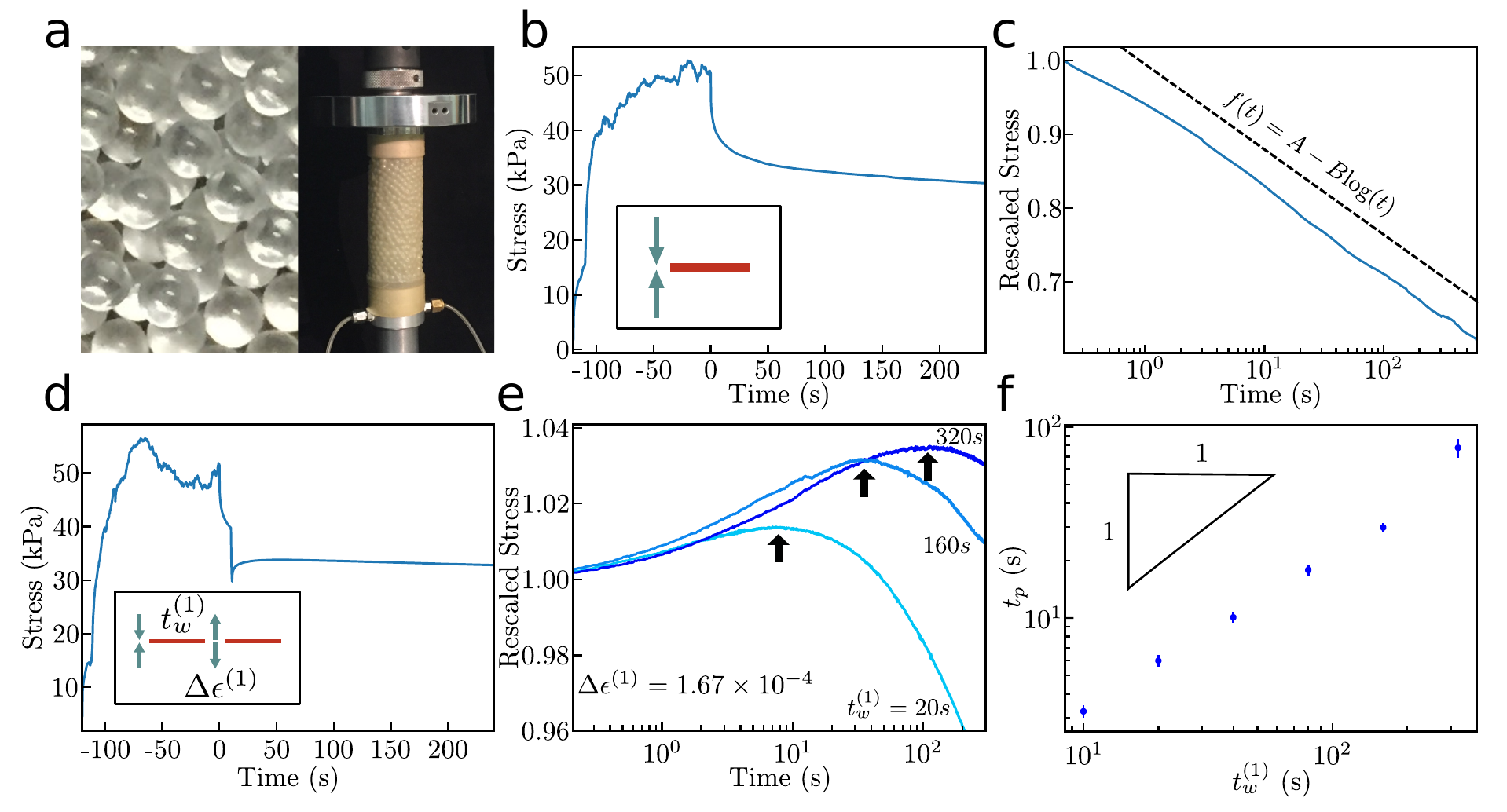}
	\caption{Anomalous relaxation in a granular packing of glass spheres. \textbf{(a)} The 5mm glass spheres used in experiment, magnified and in the latex membrane.  \textbf{(b)} Stress-time data for single step relaxation, where a packing is compressed to strain $\epsilon=0.1$ and then held statically for the remainder of the experiment. Inset illustrates the experimental protocol: arrows pointing inward (outward) correspond to compression (decompression), and red lines denote holds at constant strain.
	\textbf{(c)} The stress decays approximately logarithmically from the moment compression halts.  These data are from \textbf{(b)}, with stress normalized by its value at the start of the hold.
	\textbf{(d)} Stress-time data for two-step relaxation. 
	\textbf{(e)} The stress of two-step relaxation for three different $t_w^{(1)}$, with turnaround time marked by arrows.
	\textbf{(f)} The turnaround time, $t_p$, scales approximately linearly with $t_w^{(1)}$, for constant $\Delta\epsilon^{(1)}$.}
	\label{fig1}
\end{figure*}

For each experiment, 5mm diameter soda lime glass spheres (MoSci) were poured into a 5cm diameter latex membrane to form a 2:1 column (height to diameter) (Fig. 1a).  Isotropic confinement of 40kPa was applied via holding the interior of the membrane at lower pressure.  The column was compressed by an Instron universal materials tester to a strain of $\epsilon=0.1$, as a fraction of the initial (uncompressed) height, over two minutes.  
During this initial compression, stress builds across the packing and grain reconfigurations lead to stress fluctuations (Fig. 1b).
At $\epsilon=0.1$ compression halted and the force was recorded over several minutes at
a rate of 100 data points per second.  To program memories into the packing, the Instron was driven forward or backward by small strain amounts $\Delta\epsilon^{(i)}$ after various wait times $t_w^{(i)}$ before holding again.  The nonmonotonic stress relaxation of interest is found in the data from the final hold, during which nothing further is done to the granular system beyond monitoring the stress.

Simple uniaxial compression followed by a hold without any additional steps of (de)compression, leads to stress relaxation of the granular material that is approximately logarithmic in time (Fig. 1c), in agreement with past results \cite{MakseRelaxation2005,LogAgingAcoustic2013}.  
This logarithmic relaxation suggests that granular materials might belong in the company of systems able to exhibit also nonmonotonic relaxation after an appropriate driving protocol, as seen previously for crumpled paper \cite{Lahini2017} or the frictional interface between two blocks of PMMA \cite{Dillavou2018}. 

One such protocol is as follows: rather than holding the system at a strain state $\epsilon$ indefinitely, it is allowed to relax partially for some time $t_w^{(1)}$, but then the applied strain is \textit{decreased} to $\epsilon-\Delta\epsilon^{(1)}$. 
After driving a granular packing in this way, it relaxes in a strikingly non-intuitive manner: without any additional prompting, nor any additional energy input to the system, the 
stress measured \textit{increases} 
for a period of time before turning around and resuming a slow decrease 
that shows no signs of stopping on experimental time scales.  Further, 
the timescale $t_w^{(1)}$ of the hold in the strain state $\epsilon$ 
emerges as a memory which is revealed in the turnaround time $t_p$ (Fig. 1d,e).

We employ AOI DRT to rationalize nonmonotonic relaxation, as was 
previously done by Lahini et al. \cite{Lahini2017}.  Within the framework of DRT, a system is idealized as a population of simple, exponential relaxers in parallel with a distribution of relaxation rates $\lambda$. In the AOI variant \cite{Imry2012}, $P(\lambda)\sim1/\lambda$ over a range $[\lambda_\text{min},\lambda_\text{max}]$, yielding logarithmic relaxation over timescales 
between $\lambda_\text{max}^{-1}$ and $\lambda_\text{min}^{-1}$. Compelling reasons in support of this distribution of relaxation rates were described in earlier work on luminescence \cite{MaxEntDecay1987}. $P(\lambda)\sim1/\lambda$ is uniform in log$\lambda$ space and identical whether working in terms of rates $\lambda$ or timescales $\tau=1/\lambda$. Both $\lambda$ and $\tau$ are scale parameters -- i.e., domain of (0, $\infty$) -- and as such this distribution is the maximum entropy distribution \cite{Jaynes1968} for which minimal prior information has been assumed.  The distribution maximizes generality, providing a reason for the widespread occurrence of logarithmic relaxation.

Each of the relaxing elements holds a portion of stress $\Gamma(\lambda)$ which dissipates exponentially in time according to $d\Gamma=-\lambda\Gamma dt$, and the measured signal (i.e., the total stress) is the sum over all elements.  Compression affects all elements equally and is taken to occur over a timescale negligible to even the fastest elements.  Figure 2c shows the system state at various stages of relaxation, where the elements are displayed from slowest (left, dark) to fastest (right, light).  At a time $t_w$ into the relaxation (Fig. 2c II), the fastest elements have relaxed to $\Gamma=0$ stress and the slowest still bear most of their original stress.  

In such a state, the system has dual natures: through the slow elements it remembers the initial, unstrained state, and through the fast
ones it has adapted to the strain state $\epsilon$.  Decompression at this time decreases the magnitude of the stress in the slow elements and negatively stresses the fast elements, creating a system state $\Gamma(\lambda)$ in which subsets of the elements will relax in opposing directions.  The fast elements relax first, causing the paradoxical increase in stress over time \textit{even though all elements decrease in energy}, which scales with the square of the stress.  Eventually the slow elements turn the relaxation around, giving rise to nonmonotonic dynamics.  The memory is clearly visible in the stress held across the relaxing elements, $\Gamma(\lambda)$, in state III of Fig. 2c, where the timescale of the switch in sign was imprinted by the duration of the hold $t_w^{(1)}$.

Casting the nonmonotonic relaxation of the granular packing within the DRT framework shows that the storage capacity of memories should be 
larger than two.  Specifically, by appending a small strain step in the original (positive) direction, we can create three steps of relaxation in a simulated system (Fig. 2b).  Again, the memory is manifest in $\Gamma(\lambda)$ of state V in Figure 2c, where the population of relaxing elements has been split into three counter-relaxing contingents.

\begin{figure}
	\centering
	\includegraphics[width=1\linewidth]{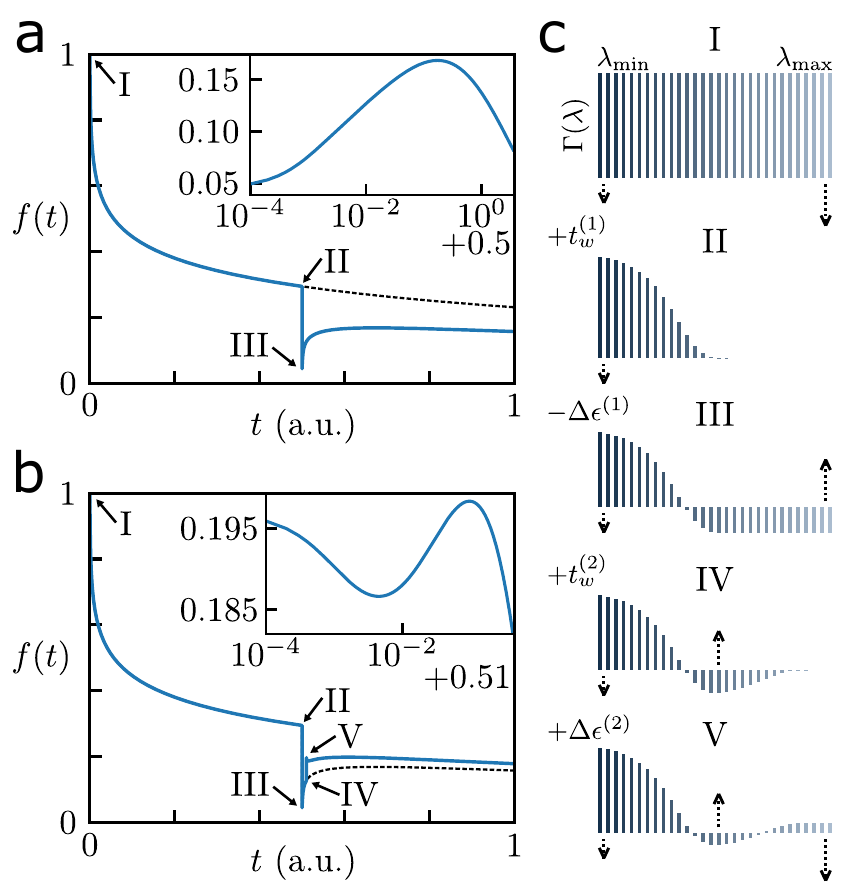}
	\caption{The Amir-Oreg-Imry variant of the different relaxation times framework (AOI DRT).
	\textbf{(a)} Simulated stress-time data for two-step relaxation, with the inset showing the relaxation from the start of the final hold.
	\textbf{(b)} Simulated stress-time data for three-step relaxation, with the inset showing the relaxation from the start of the final hold.  In \textbf{(a)} and \textbf{(b)} the dashed line shows the relaxation of $f(t)$ without the extra strain step $\Delta\epsilon^{(1)}$ and $\Delta\epsilon^{(2)}$, respectively.
	\textbf{(c)} The state of the system $\Gamma(\lambda)$ at various points in time, shown as the stress held by each element ordered from slowest (dark, left) to fastest (light, right).  The time of the states are marked in \textbf{(a)} and \textbf{(b)} as Roman numerals I-V.}
	\label{fig2}
\end{figure}

\begin{figure}
	\centering
	\includegraphics[width=0.9\linewidth]{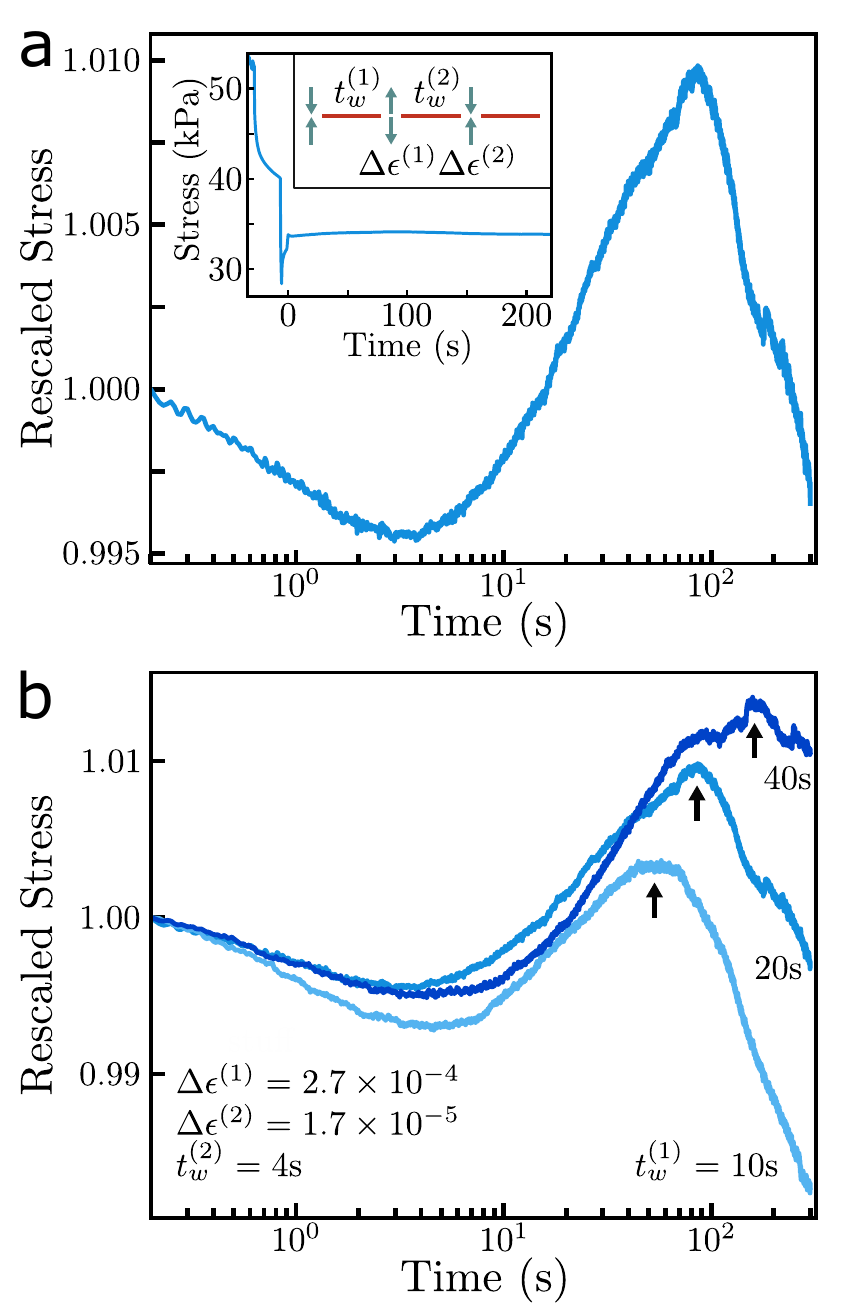}
	\caption{Three-step relaxation of glass beads. \textbf{(a)} Stress during the final hold of three-step relaxation, rescaled by its value at the start. \textit{Inset:} Stress-time data for the various stages of relaxation, showing the two wait times $t_w^{(1)}$ and $t_w^{(2)}$ before the final hold.  
	\textbf{(b)} Varying $t_w^{(1)}$ while holding the rest of the driving protocol constant shifts the time of the second turnaround in stress.  Memories are thus played out in the reverse order from which they were stored during driving.}
	\label{fig3}
\end{figure}

Guided by the simulated system, we find three-step relaxation -- the first observed in any disordered system, to the best of the authors' knowledge -- in our packings
by adding a small compression step $\Delta\epsilon^{(2)}$ in the forward direction, after waiting time $t_{w}^{(1)}$ at $\epsilon$ and then $t_{w}^{(2)}$ at $\epsilon-\Delta\epsilon^{(1)}$ (Fig. 3a).  The resulting stress during relaxation undulates back and forth without any intervention: it
decreases, increases, and then decreases again over timescales imprinted during the loading process.  
After fixing the strain state for the final hold, nothing is done to the granular system to prompt the nonmonotonicity; thus at the start of the hold the packing is in a state that `knows' to turn around in stress after 4 seconds and then again some 90 seconds later.  

The AOI DRT model also predicts a functional form for the relaxation: a sum of alternating logarithms, staggered in time according to the start of the prior strain steps (see Figs. 4a, b).  
\begin{figure}
	\centering
	\includegraphics[width=1\linewidth]{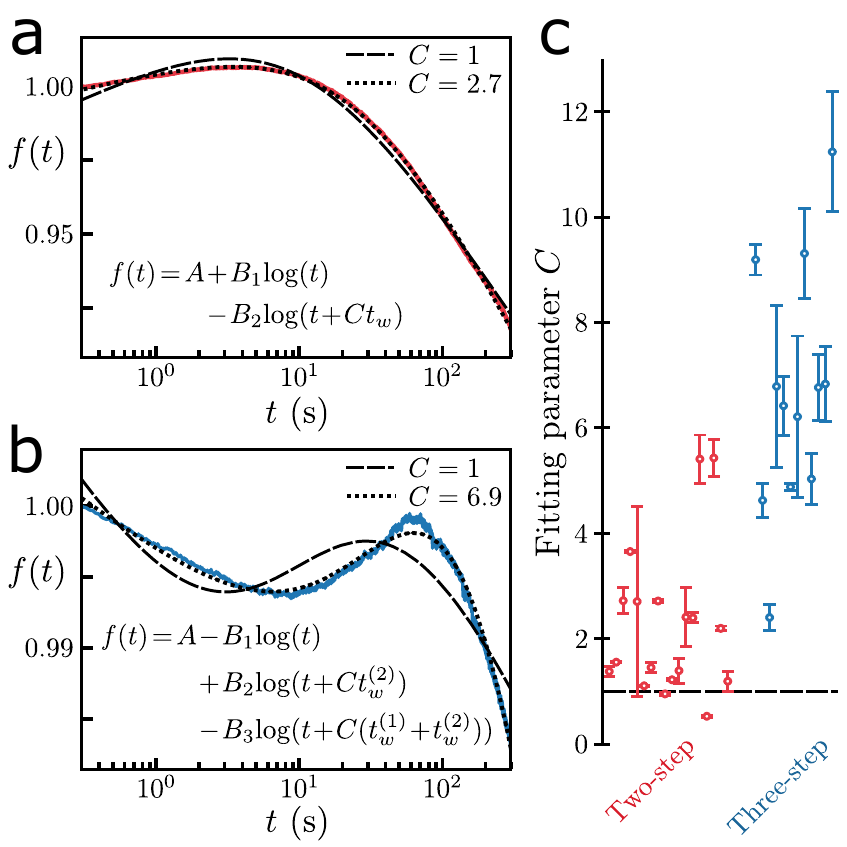}
	\caption{Fitting multistep relaxation data to the form predicted by the model. \textbf{(a)} An example of two-step relaxation data fitted to the form predicted by the model ($C=1$) and to a form where $C$ is allowed to vary. 
	\textbf{(b)} The same, for an example of three-step relaxation.
	\textbf{(c)} The best-fit values of the parameter $C$ across all two- and three-step relaxation experiments. The dashed line, $C=1$, represents AOI DRT.}
	\label{fig4}
\end{figure}
However, we find a correction is necessary to achieve reasonable fits to the two- and three-step relaxation data
(Fig. 4).  
Specifically, we find that the wait times in the logarithms have to be multiplied by a parameter $C >$ 1.  This 
does not arise from the model (where $C$ = 1) and indicates a deficiency in its descriptive power.  
$C >$ 1 was also necessary for properly fitting the two-step relaxation in Ref. \cite{Lahini2017} though no attention was called to it, presumably because $C$ was still close to unity.  We find that three-step relaxation data significantly increases the discrepancy between the AOI model and experimental results, with fitted $C$ values often an order of magnitude larger than the model allows (Fig. 4c).

The implications reach beyond granular physics to the wide range of disordered systems displaying anomalous relaxation.  The AOI DRT model is valuable for its simplicity and generality, but the data of Fig. 4 show that it fails to capture a key aspect of the physics.  

\begin{figure}
	\centering
	\includegraphics[width=1\linewidth]{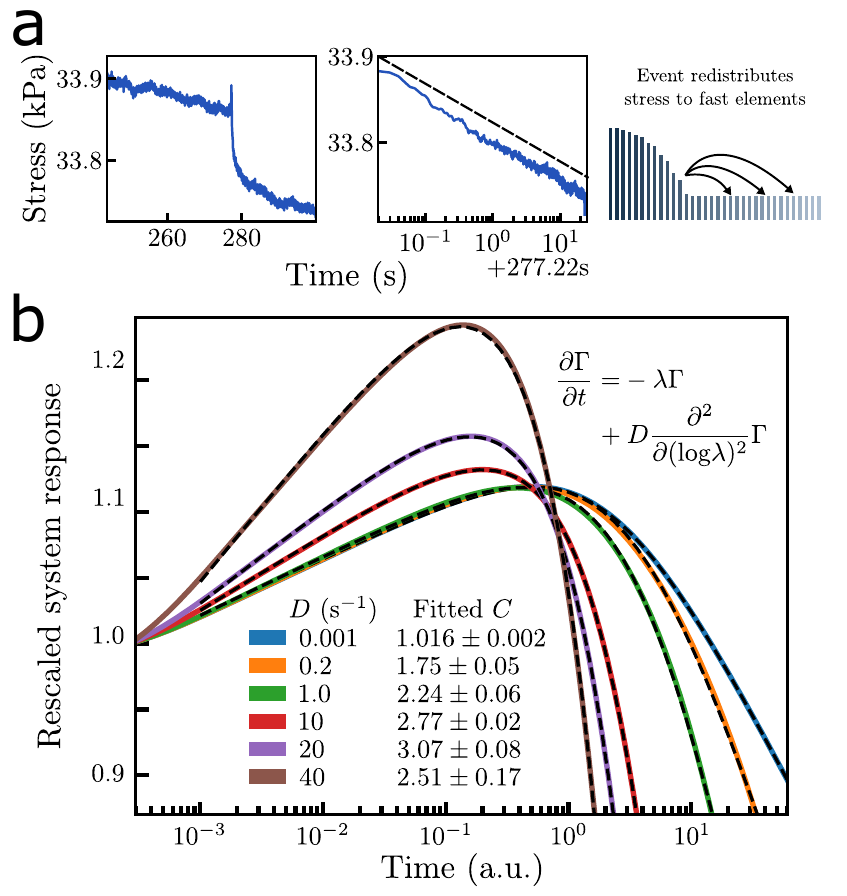}
	\caption{Crosstalk between elements in the model. \textbf{(a)} A sudden relaxation event late into one of the experiments (left panel, at $t$ =277.22s) leads to a renewal of logarithmic decay in stress (right panel).  In the framework of DRT, this occurs as a renewal of the fastest relaxing elements.
	\textbf{(b)} Simulated two-step relaxation where the driving protocol is held constant and diffusion between elements is varied.  The fitting function is the same as in Fig. 4a.}
	\label{fig5}
\end{figure}

We conjecture about $C >$ 1 based on observations of discrete events that occur occasionally during the relaxation of a granular packing.  During such events, the stress or its derivative suddenly change in magnitude.  In many cases, one of which is shown in Fig. 5a, an event renews fast timescale relaxation late into a static hold.  In terms of  AOI DRT, the event effectively ``rejuvenated'' faster relaxing elements which had long since adapted to the current strain state.  This is suggestive of crosstalk between the relaxing elements, occurring in discrete steps during events such as the one shown and ostensibly continuously during the gradual transformation of the granular packing.  That the renewed relaxation is approximately logarithmic over more than two orders of magnitude in time shows the crosstalk redistributed stress to the fast elements nearly uniformly. 

A simple form of crosstalk between elements can be incorporated into the model through an effective diffusion of stress in $\Gamma(\lambda)$.  The diffusion is included as a Laplacian in $\text{log} \lambda$ space, scaled by a coefficient $D$.  Implemented in this way, all relaxation curves (Fig. 5b) are fit by the same series of logarithms as in Fig. 4a, and the fitting parameter $C$ grows from 1 in the absence of diffusion to as large as 3 for the specific ($t_w^{(1)}$, $\Delta\epsilon^{(1)}$) two-step protocol.  With faster stress diffusion, the value of $C$ decreases until eventually the nonmonotonicity vanishes.

The diffusion of stress between relaxers lies in the same vein as modifications to DRT where the relaxation timescales can evolve in time \cite{NMRRateDiffusion1995}.  It offers a means to incorporate coupling between elements and to venture into the gulf between parallel and sequential models of relaxation (e.g., \cite{PalmerHierarchical1984,Klafter3Relaxers1986,BreyHierLog2001}).  However, we also find that this implementation of diffusion modifies single step relaxation away from logarithmic at large timescales, and preliminary sweeps of three-step relaxation did not lead to values of $C$ much greater than 2.  

In this work, the seemingly paradoxical behavior of a relaxing disordered system has been investigated in a new guise: a granular packing of glass spheres.  The general AOI DRT model was used to explain two-step relaxation and predict three-step relaxation before exposing its own deficiency in the fitting function for the nonmonotonic relaxation.  It is expected that the understanding derived from this case study translates to the myriad systems exhibiting anomalous relaxation to a perturbation.

\textbf{Acknowledgements:} We thank Tom Witten, Sidney Nagel, Shmuel Rubinstein, and Thomas Videb{\ae}k for insightful discussions.  This work was supported by the NSF through through grant CBET-1605075 and by the Chicago MRSEC through grant DMR-1420709.
K.M. acknowledges support from the Center for Hierarchical Materials Design (CHiMaD).

\end{document}